\documentclass[a4paper]{article}

\usepackage{INTERSPEECH2022}
\usepackage[nolist]{acronym}

\usepackage{float}
\usepackage{subcaption}

\usepackage{mathtools}
\usepackage{bm}

\DeclarePairedDelimiter\paren{(}{)}           
\DeclarePairedDelimiter\abs{\lvert}{\rvert}   
\DeclarePairedDelimiter\norm{\lVert}{\rVert}  
\DeclarePairedDelimiter\bkt{[}{]}             
\DeclarePairedDelimiter\set{\{}{\}}           

\newcommand{\mvec}[1]{\boldsymbol{#1}} 
\newcommand{\mmat}[1]{\boldsymbol{#1}} 

\usepackage{adjustbox}
\usepackage{makecell}

\title{Time-Domain Based Embeddings for Spoofed Audio Representation}
\name{Matan Karo$^1$, Arie Yeredor$^1$, Itshak Lapidot$^{2,3}$}
\address{
  $^1$Tel Aviv University (TAU), School of Electrical Engineering, Israel\\
  $^2$Afeka Tel Aviv Academic College of Engineering, ACLP, Israel \\
  $^3$Avignon University, LIA, France}
\email{matankaro@mail.tau.ac.il, itshakl@afeka.ac.il, arie@eng.tau.ac.il}

\begin{document}
\begin{acronym}[CFCCIF]
	\newacro{APGF}{All-Pole Gammatone Filter} 
	\newacro{ASV}{Automatic Speaker Verification} 
	\newacro{AUC}{Area Under Curve}
	\newacro{CC}{Cepstral Coefficients} 
	\newacro{CDF}{Cumulative Probability Function} 
	\newacro{CQCC}{Constant-Q Cepstral Coefficients} 
	\newacro{CFCCIF}{Cochlear Filter Cepstral Coefficients with Instantaneous Frequency}
	\newacro{DET}{Detection Error Tradeoff} 
	\newacro{DM}{Diffusion Maps} 
	\newacro{DNN}{Deep Neural Network} 
	\newacro{DR}{Dimensionality Reduction} 
	\newacro{DTW}{Dynamic Time Warping} 
	\newacro{EM}{Expectation Minimization}
	\newacro{EER}{Equal Error Rate} 
	\newacro{ERB}{Equivalent Rectangular Bandwidth} 
	\newacro{ERR}{Equal Error Rate} 
	\newacro{FAR}{False Acceptance Rate} 
	\newacro{FN}{False Negative} 
	\newacro{FNR}{False Negative Rate} 
	\newacro{FP}{False Positive} 
	\newacro{FPR}{False Positive Rate} 
	\newacro{FRR}{False Rejection Rate} 
	\newacro{GAN}{Generative Adversarial Network} 
	\newacro{GCF}{Gammachrip Filter} 
	\newacro{GMM}{Gaussian Mixture Model} 
	\newacro{GMM-UBM}{Gaussian Mixture Model Universal Background Model}
	\newacro{GTCC}{Gammatone Cepstral Coefficients} 
	\newacro{GUI}{Graphical User Interface} 
	\newacro{HLL}{Human Log-likelihoods} 
	\newacro{HMM}{Hidden Markov Model} 
	\newacro{IGTCC}{Inverted Gammatone Cepstral Coefficients} 
	\newacro{IMFCC}{Inverted Mel Frequency Cepstral Coefficients} 
	\newacro{IoT}{Internet of Things} 
	\newacro{JDGMM}{Joint Density Gaussian Mixture Model} 
	\newacro{JSD}{Jensen-Shannon Divergence} 
	\newacro{KLD}{Kullback-Leibler Divergence} 
	\newacro{KLS}{KL-Symmetric} 
	\newacro{LA}{Logical Access} 
	\newacro{LFCC}{Linear Cepstral Coefficients} 
	\newacro{LR}{Logistic Regression} 
	\newacro{LSTM}{Long Short-Term Memory} 
	\newacro{LTI}{Linear Time-Invariant} 
	\newacro{MDS}{Multidimensional Scaling} 
	\newacro{MFCC}{Mel Frequency Cepstral Coefficients} 
	\newacro{MGDF}{Modified Group Delay Function} 
	\newacro{NLP}{Natural Language Processing} 
	\newacro{NN}{Neural Network} 
	\newacro{OOS}{Out Of Sample} 
	\newacro{OZGF}{One-Zero Gammatone Filter}
	\newacro{PA}{Physical Access} 
	\newacro{PCA}{Principal Component Analysis} 
	\newacro{PDF}{Probability Density Function} 
	\newacro{PMF}{Probability Mass Function} 
	\newacro{RNN}{Recurrent Neural Network} 
	\newacro{ROC}{Receiver Operating Characteristic} 
	\newacro{SAS}{Spoofing and Anti-Spoofing} 
	\newacro{SD}{Speech Deepfake} 
	\newacro{SI}{Speaker Identification} 
	\newacro{SPSS}{Statistical Parametric Speech Synthesis} 
	\newacro{SS}{Speech Synthesis} 
	\newacro{STD}{Standard Deviation} 
	\newacro{SV}{Speaker Verification} %
	\newacro{SVM}{Support Vector Machine} 
	\newacro{TN}{True Negative} 
	\newacro{TNR}{True Negative Rate} 
	\newacro{TP}{True Positive} 
	\newacro{TPR}{True Positive Rate} 
	\newacro{TTS}{Text-to-Speech} 
	\newacro{VAD}{Voice Activity Detection} 
	\newacro{VC}{Voice Conversion} 
\end{acronym}

\maketitle
\begin{abstract}
Anti-spoofing is the task of speech authentication. That is, identifying genuine human speech compared to spoofed speech.
The main focus of this paper is to suggest new representations for genuine and spoofed speech, based on the probability mass function (PMF) estimation of the audio waveforms' amplitude.

We introduce a new feature extraction method for speech audio signals: unlike traditional methods, our method is based on direct processing of time-domain audio samples. The PMF is utilized by designing a feature extractor based on different PMF distances and similarity measures. As an additional step, we used filter-bank preprocessing, which significantly affects the discriminative characteristics of the features and facilitates convenient visualization of possible clustering of spoofing attacks. Furthermore, we use diffusion maps to reveal the underlying manifold on which the data lies.

The suggested embeddings allow the use of simple linear separators to achieve decent performance.
In addition, we present a convenient way to visualize the data, which helps to assess the efficiency of different spoofing techniques.
  
The experimental results show the potential of using multi-channel PMF based features for the anti-spoofing task, in addition to the benefits of using diffusion maps both as an analysis tool and as an embedding tool.
  \end{abstract}

\section{Introduction}

Biometrics technologies have a major role in personal identification and authentication,
voice-based systems in particular~\cite{Biometrics2019}.
The purpose of an \ac{ASV} system is to validate the speaker's identity by analyzing an audio sample of his/her voice~\cite{ASV_main,hansenSpeakerRecognitionMachines2015}.
ASV systems allow the user to authenticate its identity in a natural, non-invasive manner. As the use of ASV systems is becoming more popular, the interest in bypassing and exploiting them increases accordingly.
Therefore, the issue of preventing attackers from exploiting such systems is necessary.

The counter task of speech spoofing, hence detecting the fraud, is called Anti-Spoofing, and it has been studied vastly as described, e.g., in selected surveys~\cite{wuSpoofingCountermeasuresSpeaker2015,wuAntiSpoofingTextIndependentSpeaker2016}.
Due to the importance of finding countermeasures, a series of anti-spoofing challenges has been published~\cite{ASVspoofWebsite}.
Numerous of studies have been made using the challenges databases, and comparisons between different methods can be found in~\cite{kambleAdvancesAntispoofingPerspective2020a,nautschASVspoof2019Spoofing2021,2019_review}.
Our work has been done primarily on the ASVspoof 2019~\cite{ASVspoof2019} \ac{LA} database.

Most speech features for anti-spoofing are based on frequency-domain analysis.
There is almost no research directly addressing time-domain based features as spoofing countermeasures. However, the time-domain representation contains information which is utilized for various tasks~\cite{time_features_4}.
Previous work~\cite{itzik_1,itzik_2} has shown that the \ac{PMF} of genuine speech samples differs from the PMF of spoofed speech samples.
Motivated by those results, our goal is to further expand such PMF-based methods by exploring the use of channel-based time-domain features.
For exploring spoofed speech from a time-domain perspective, we propose a new feature extraction approach, which relies on estimating the samples' joint or marginal PMFs with the addition of filter-bank pre-processing.

In previous work, the probability mass function, which is equivalent to the normalized histogram of the audio samples' amplitude, was calculated across each set of the database as a whole.
To get a better understanding of the separative characteristics of the speech PMF for the classification task, we have utilized the PMF in a different way, using it in the feature extraction process.
First, gender-based labels have been added to the audio files under the assumption that the speaker's gender influences speech distribution~\cite{gender_pmf}.
Due to the pre-known target's gender in the ASV task, gender separation is a valid assumption.
The effect of gender separation is discussed in Section~\ref{sec:experiments}.

The feature extractor included two global speakers models, genuine and spoofed, which were calculated as the PMF across the training set.
Then, similarity measures (such as distance metrics) from each model were calculated for producing a single feature, corresponding to a specific measure. This process is given in further details in Section~\ref{sec:feature_extraction}.

Frequently used features for the anti-spoofing are cepstral features~\cite{sahidullahComparisonFeaturesSynthetic2015}, such as \ac{MFCC}, \ac{LFCC} and \ac{CQCC}~\cite{lfcc_vs_mfcc, CQCC}.
Inspired by the feature extraction process of spectral coefficients, different filter-banks were applied to the audio samples.
Each filter output created an output channel, for which the PMF distances from each global speaker model were calculated. The motivation was to identify dominant channels for the anti-spoofing task.

The main research objective was to gain new insights about genuine and spoofed speech and prominent differences between these two in the respective samples' PMFs.
Due to the difficulty to distinguish between them, we have looked for new ways to embed the relations between genuine and spoofed speech, and between different attack methods.
Therefore, this work's main focus is on feature representation other than optimizing a state-of-the-art classification scheme.
For that purpose, we have used a manifold learning technique known as \ac{DM}~\cite{dm_2}, for both visualization and features embedding.
The use of DM revealed the discriminative quantities of the PMF method and provided new insights about the effectiveness of different attacks. A simple \ac{LR} was used for classification.

\section{PMF-based Feature Extraction}
\label{sec:feature_extraction}

\begin{figure}
    \centering
    \includegraphics[width=\linewidth]{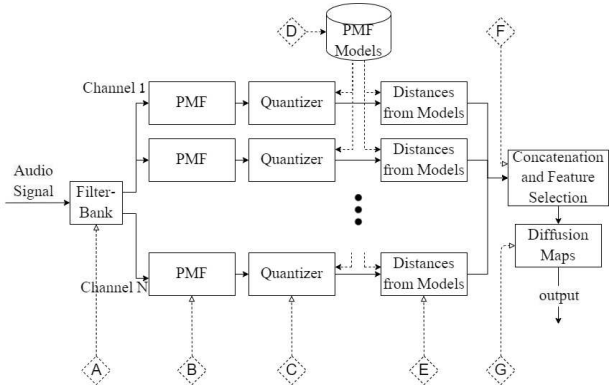}
    \caption[Features extraction scheme]{Features extraction scheme.}
    \label{fig:feature_extraction_scheme}
\end{figure}

This section describes the feature extraction process, or in other words, the embedding of human and spoofed speech utterances into a new representation.
The core idea is that spoofed speech affects the amplitude's distribution in a way that will cause a distinguishable difference between human and spoofed speech.
In both~\cite{itzik_1} and~\cite{itzik_2} PMFs of different datasets were calculated, showing differences between human and spoofed distributions.
In this work, we have utilized the PMFs as a part of a feature extraction scheme whose illustration is given in Figure~\ref{fig:feature_extraction_scheme}.

\subsection{Filter-Banks}
\label{ssec:filter_banks}

The design process of filter-bank in speech processing is frequently influenced by a biological or psychoacoustic model of the human hearing system/perception.
Due to the use of such filters in anti-spoofing solutions, we chose to include them in our scheme, but with some integration modifications.
To begin with, our PMF-based approach is meant to examine the samples in the time-domain instead of the frequency-domain, as is normally the case with filter-banks.
Therefore, we applied the filtering process directly on the audio samples.


We chose to focus on two filter-banks types: Mel filter-bank (used in MFCC~\cite{si_3}) and Gammatone filter-bank (used in \ac{GTCC}~\cite{valeroGammatoneCepstralCoefficients2012}):
\begin{itemize}
    \item \underline{Mel Filter-Bank}: a set of triangular, variable bandwidths filters, spaced linearly on mel scale. The design simulates the non-linear frequeuncy-sensitivity of the human hearing system - its resolution degrades as the frequency increases.

    \item \underline{Gammatone Filter-Bank}: a set of gammatone filters, equally spaced on the ERB scale~\cite{GammatoneBase2}. Gammatone filter banks were designed to model the human auditory system. Inspired by the cochlea structure, a gammatone filter-bank simulates its action by transforming complex sounds into a multichannel activity pattern, in a similar manner to the pattern observed in the auditory nerve.
\end{itemize}

Although the use of gammatone filter-banks not as common as the use of mel and linear filter-banks in traditional anti-spoofing solutions, gammatone based features have proven themselves both in speech recognition~\cite{kimPowerNormalizedCepstralCoefficients2016}
and in anti-spoofing tasks~\cite{valeroGammatoneCepstralCoefficients2012,Phase3_mel_gm}.

In this paper we will focus on the Gammatone filter-bank, which showed superior separative capabilities compared to the Mel filter-bank. Furthermore, we have also used an inverted version of the Gammatone filter-bank, known as Inverse Gammatone filter-bank, which enabled to separate specific spoofing methods.

\subsection{Probability Mass Function}
\label{ssec:probability_mass_function}

The probability mass function (PMF) is a function which represents the probabilities of a discrete random variable to attain each of its possible values..
In our context, we will refer to the PMF as the normalized histogram of a discrete set of attainable values.
Analog audio waveform representation is continuous, meaning its statistical representation should be done by using a continuous function such as \ac{PDF}.
However, in this work's framework the audio samples are digital, hence quantized to a finite set of values. Therefore, the PMF is used and approximates the PDF of the real analog audio.
This approximation depends on the bin width $\Delta x$, which is the width of each bin in the histogram. The $PMF$ of bin $x_0$ can be approximated by:
\begin{align}
    PMF\paren*{x=x_{0}} \approx \int_{x_{0}-\frac{\Delta x}{2}}^{x_{0}+\frac{\Delta x}{2}}PDF\paren*{x}\,dx
\end{align}
For waveform samples which are sampled using $b$ bits per sample, there are $2^b$ possible discrete values.
For both ASVspoof 2015 and ASVspoof 2019 the LA audio files were sampled with 16 bits per sample and normalized to $\bkt*{-1,1}$, which dictates explicitly the histogram bins for the direct PMF calculation.
The PMF calculation for filtered audio was done after inspecting the dynamic range of the output channels of the training set. After the inspection we chose to keep the bins the same and to clip exceeding values.

\subsection{Similarity Measures}
\label{ssec:similarity_measures}

Inspired by the \ac{GMM-UBM}~\cite{gmm-ubm} approach, which uses a global speaker model for the feature embedding process,
we have also taken a global model-based approach. The PMFs calculated in~\cite{itzik_2} are utilized to represent a global human or spoofed speaker model.
Models for human and spoofed speech have been constructed by calculating the PMF across corresponding audio files in the training set.
Then, each audio file can be represented by a similarity measure to each model. To this end, a PMF similarity measure (such as distance metric), should be defined~\cite{general_hist_dist1}.

In~\cite{itzik_1} and~\cite{itzik_2}, \ac{KLD}~\cite{kl_dist_and_div} was discussed as an evaluation metric for the similarity between PMFs.
In our work, the use of \ac{KLD} as a similarity measure was expended to using several similarity measures: quadratic Chi distance~\cite{chi_dist}, normalized cross correlation~\cite{normed_corr}, Hellinger distance~\cite{hellinger_distance}, histogram intersection~\cite{histogram_intersection_dist}, Jensen-Shannon~\cite{js_dist}, Symmetrized Kullback-Leibler and Kullback-Leibler Divergence~\cite{kl_dist_and_div}, and modified Kolmogorov-Smirnov~\cite{ks2_modified}.

\subsection{Features Concatenation and Selection}
Feature construction is the process of generating a single feature vector out of the extracted features. 
In our case, each extracted feature expresses the similarity of an input PMF to a model PMF under a certain measure for a specific channel.
We have tested several concatenation schemes, some isolate each channel and distance combination, some concatenating groups of features and some more sophisticated methods which involved pre processing before the concatenation.
These schemes are described and compared in Section~\ref{sec:experiments}.

\section{Diffusion Maps}
\label{sec:diffusion maps}

Diffusion maps (DM) is a part of nonlinear dimensionality reduction methods called manifold learning, which is often also used for feature extraction.
The main underlying assumption in such methods is that the high-dimensional sampled data lies on some low-dimension manifold in the high-dimensional ambient space. By utilizing the manifold structure, a reduced representation can be found, while preserving the local relations between the data samples.
Diffusion maps technique has been used successfully
in machine learning and data mining applications, including speech processing tasks~\cite{dm_example_speech1,dm_example_speech2}.

DM have an advantage over several other methods, such as the popular t-distributed stochastic neighbor embedding (t-sne)~\cite{tsne}, due to its ability to conveniently expand its embedding to new samples.
The out of sample extension ability is important in our case, due to the requirement to evaluate new data which may contain both previously observed and previously unobserved attacks.

Diffusion maps can be defined by three steps: connectivity, diffusion distance and diffusion maps embedding~\cite{porteIntroductionDiffusionMaps2008}.

\subsection{Connectivity}
\label{sec: dm_connectivity}
Let $\mmat{X} = \set*{x_1, ..., x_N} \in \mathbb{R}^d$ be a $d$-dimensional set of N data-points.


The one-step transition probability from node $x_i$ to $x_j$ can be expressed by normalizing a kernel function asymmetrically by the nodes' degree:
\begin{align}
\label{eq:dm_p_as_kernel}
    p(x_i,x_j) = \frac{k_\epsilon(x_i,x_j)}{\mathrm{deg}(x_i)}
\end{align}
where $k_\epsilon(x_i,x_j)$ is a kernel function, usually chosen as a Gaussian kernel with a width controlled by $\epsilon$, and $\mathrm{deg}(x) = \sum_{y \in \mmat{X}}{k_\epsilon(x,y)}$.
The elements of the transition matrix $\mmat{P}$ are given by Equation~\ref{eq:dm_p_as_kernel}, and its bi-orthogonal right and left eigenvectors are denoted as $\psi_k$ and $\phi_k$, with a sequence of eigenvalues: $\abs*{\lambda_0} = 1 \ge \abs*{\lambda_1} \ge ... \ge \abs*{\lambda_{d-1}}$.


\subsection{Diffusion Distance}

The diffusion distance measures the similarity of two points in the observation space, which reflects the intrinsic geometry of the underlying manifold.
The diffusion distance at time $t$ is calculated using the connectivity, e.g. the total transition probability of all connecting paths, which describes the $t$-step evolution of the probability distribution in a Markov chain.
The diffusion distance can be expressed in terms of the eigenvectors of $\mmat{P}$:
\begin{align}
\label{eq: dm_dist_as_eigenvecs}
    D_t(x_i,x_j)^2 = \sum_{k = 1}^{d-1} \lambda_k^{2t}\paren*{\psi_k(x_i)-\psi_k(x_j)}^2
\end{align}
where $\psi_0 = \mvec{1}$ is excluded from the summation.

\subsection{Diffusion Maps Embedding}
The dimensionality reduction is achieved by approximating the diffusion distance using a lower number of summation arguments.
This is possible due to spectrum decay, which enables to approximate Equation~\ref{eq: dm_dist_as_eigenvecs} with a sufficient accuracy only by using the $K << d - 1$ first non constants eigenvectors:
\begin{align}
\label{eq: dm_dist_reduced}
    D_t(x_i,x_j)^2 \approx \sum_{k = 1}^{K} \lambda_k^{2t}\paren*{\psi_k(x_i)-\psi_k(x_j)}^2
\end{align}

\noindent Based on Equation~\ref{eq: dm_dist_reduced}, we can define the following mapping:
\begin{align}
    \Psi_t: x \rightarrow \paren*{\lambda_1^t \psi_1(x), \lambda_2^t \psi_2(x), ..., \lambda_K^t \psi_K(x)}^\top
\end{align}
The mapping $\Psi_t$ embeds the dataset $\mmat{X}$ into the Euclidean space $\mathbb{R}^K$, hence reducing the original dimension $d$ to $K$.
The diffusion distance between two points in the original dataset is equivalent to the Euclidean distance between their embeddings~\cite{dm_dist_proof}:
\begin{align}
    D_t(x_i,x_j)^2 \approx \norm*{\Psi_t(x_i) - \Psi_t(x_j)}^2
\end{align}

\subsection{Out of Sample Extension}
For large datasets the diffusion map embedding becomes impractical, as a result of oversized kernel matrix. A solution for this is to construct the embedding scheme using a subset of the complete dataset, and then apply the diffusion map to the entire dataset using an \ac{OOS} extension method.
In our framework we have chosen to use the most popular method, known as the Nystr\"{o}m's extension method, due to its simplicity and well proven performance~\cite{nystrom_oos_2}.
Given a new point $x' \notin \mmat{X}$ the eigenvector $\psi_k$ is extended to this point as follows:
\begin{align}
    \hat{\psi}_k(x') = \frac{1}{\lambda_k} \sum_{y \in \mmat{X}} p(x',y)\psi_k(y),  k = 1, ...,K 
\end{align}

\section{Experiments}
\label{sec:experiments}

As mentioned before, the main goal of our work was to gain new insights by inspecting spoofed and genuine speech in time-domain perspective. Therefore, we did not try to optimize a classifier for the embeddings, but rather focused on the embeddings representation. However, we have used a simple logistic regression classifier to quantify and compare different embeddings in terms of error rates and~\ac{EER}.

In the following experiments we have used Gammatone and Inverse Gammatone filter-banks, each with 10 channels spanned from 0$\mathrm{kHz}$ to 8$\mathrm{kHz}$. Unless mentioned otherwise, the experiments were conducted on the ASVspoof2019 database. The global speakers' PMFs were calculated per gender unless specified otherwise. In any case of DM use, it was trained using 7000 samples from the training set, 1000 from each spoofing attack and 1000 genuine, and $t$ was set to 1.

We have created a single feature vector by concatenation in the following manner:
All eight similarity measures under a given channel were concatenated by the order described in Section~\ref{sec:feature_extraction},
followed by additional concatenation of the length-8 blocks by an ascending channel order.
This process repeated itself twice, for both genuine and spoofed models, and the resulting features vectors were subtracted from each other for creating a single feature vector.

Denote $d_l^{(i)}(p_1,p_2)$ as the $l^{th}$ similarity measure between two PMFs, $p1$ and $p2$, under the $i^{th}$ channel of an arbitrary filter-bank.
Given the PMF of an input file $p_{input}$, the feature vector based on $I$ channels can be described as:
\begin{align}
\label{eq:feature_vec_with_substract}
    \mvec{f} = \paren*{s_1^{(1)}, s_2^{(1)}, ..., s_8^{(1)}, s_1^{(2)}, ..., s_8^{(I)}}
\end{align}
where $s_l^{(i)} = d_l^{(i)}(p_{input},p_{spoofed}) - d_l^{(i)}(p_{input},p_{genuine})$ and $p_{spoofed}, p_{genuine}$ are the PMFs models mentioned in Section~\ref{ssec:probability_mass_function}.
The feature vector introduced in Equation~\ref{eq:feature_vec_with_substract} is referred as "full filter-bank feature vector".
The "full feature vector" is created by concatenation of two "full filter-bank feature vectors", corresponding to the Gammatone and Inverse Gammatone filter-banks.
Other features representations may be created by taking only partial channels or distance metrics. This is done by omitting the corresponding features from the "full feature vector".


\subsection{Full Feature Embeddings}
\label{ssec:full_feature_embeddings}
\begin{figure}
    \centering
    \begin{subfigure}[b]{0.45\linewidth}
        \centering
        \includegraphics[width=\textwidth]{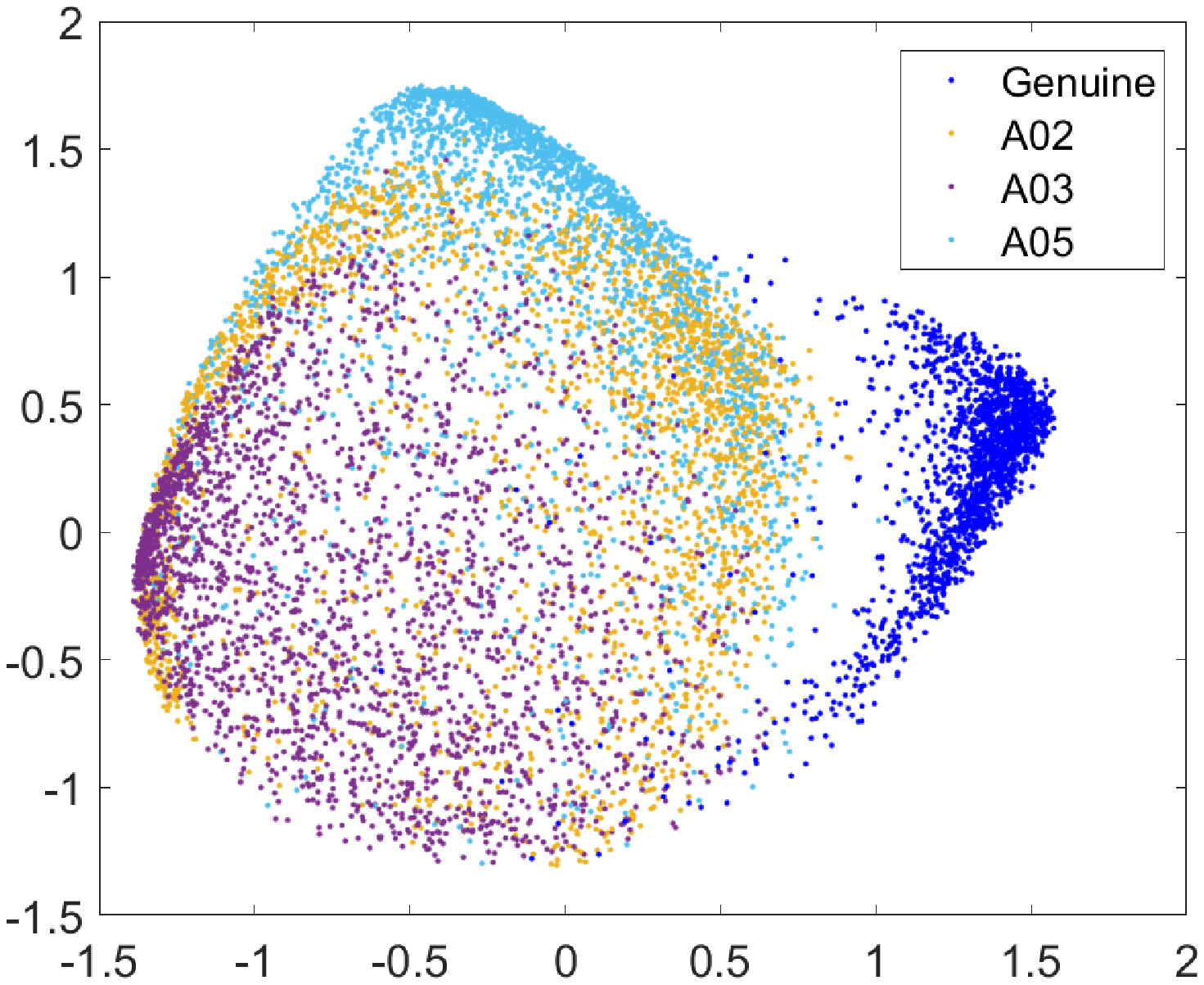}
        \caption{Gammatone with 10 channels}
        \label{fig:gm10_dm_all_dists_and_metrics_sub_and_norm_dev_female}
    \end{subfigure}
    \hfill
    \begin{subfigure}[b]{0.45\linewidth}
        \centering
        \includegraphics[width=\textwidth]{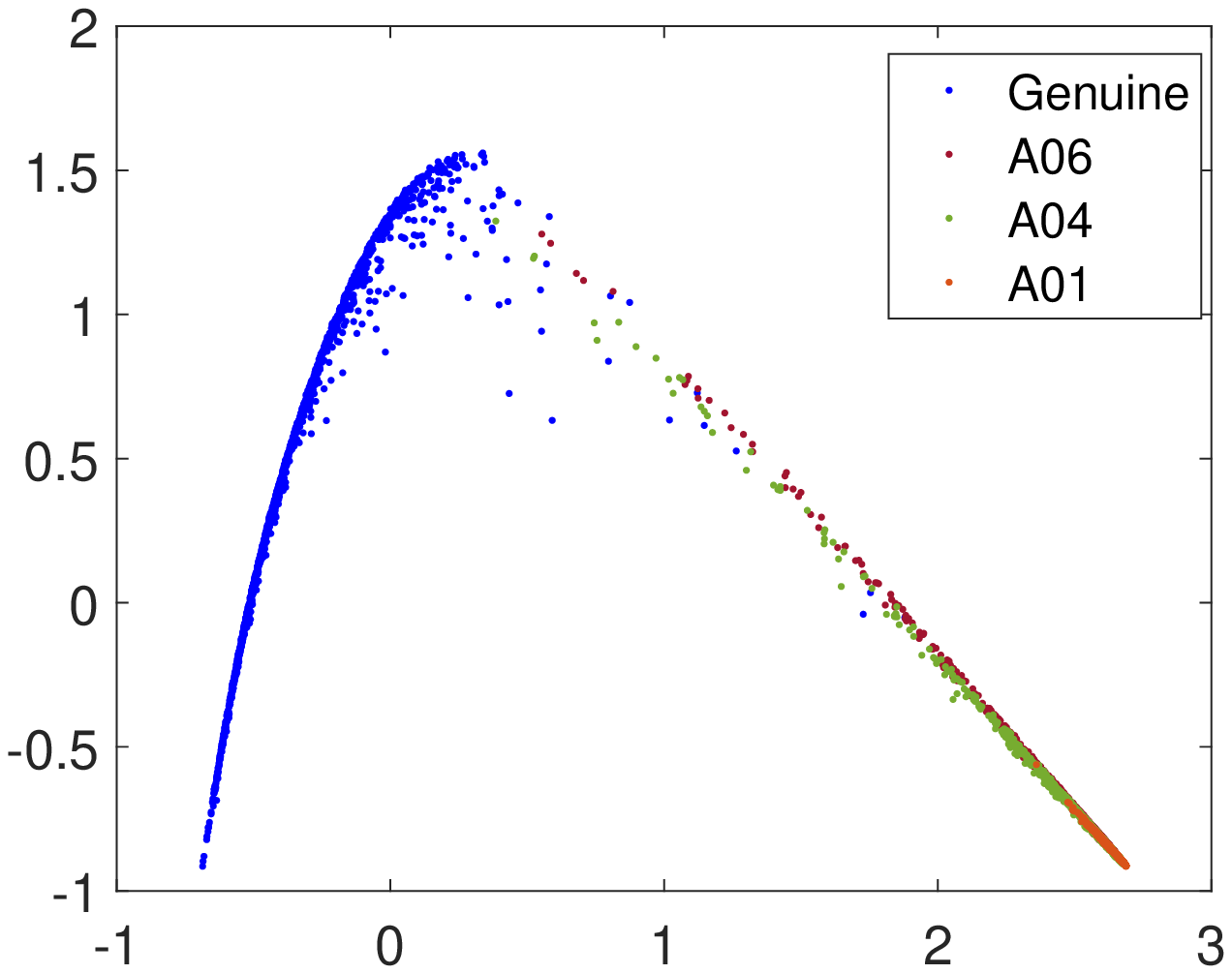}
        \caption{Inverse Gammatone with 10 channels}
        \label{fig:gm10_dm_all_dists_and_metrics_sub_and_norm_dev_female}
    \end{subfigure}    
    \caption[Gammatone vs. Inverted Gammatone DM embedding for all channels and metrics.]
    {Gammatone vs. Inverted Gammatone DM embedding for all channels and metrics. The embeddings are done on all female samples of development set.}
    \label{fig:gm10_vs_gmInv10_dm_all_dists_and_metrics_sub_and_norm_dev_female}
\end{figure}

\begin{figure}
    \centering
    \begin{subfigure}[b]{0.45\linewidth}
        \centering
        \includegraphics[width=\textwidth]{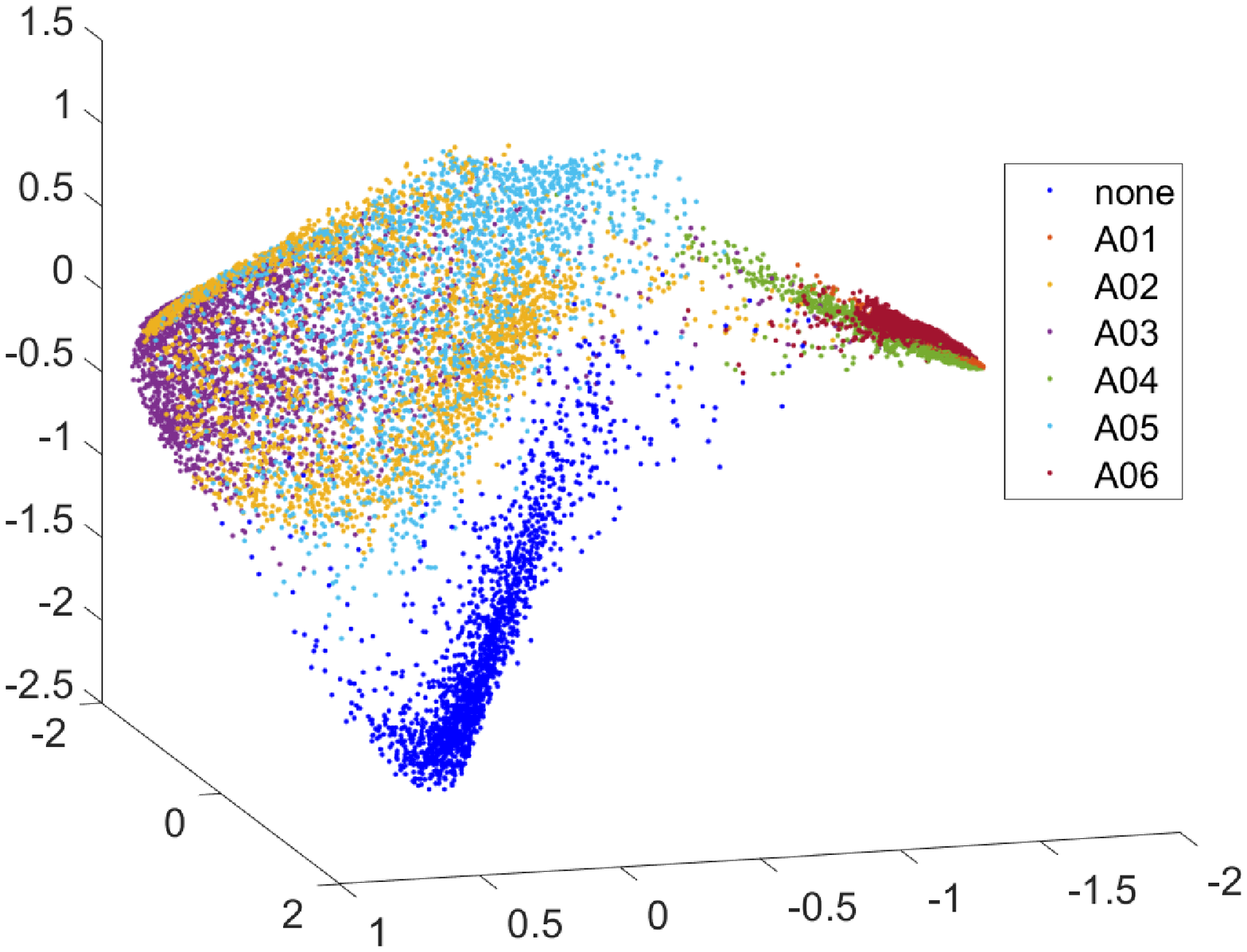}
        \caption{development set}
        \label{fig:allChAndDist_DM_female_train}
    \end{subfigure}
    \hfill
    \begin{subfigure}[b]{0.45\linewidth}
        \centering
        \includegraphics[width=\textwidth]{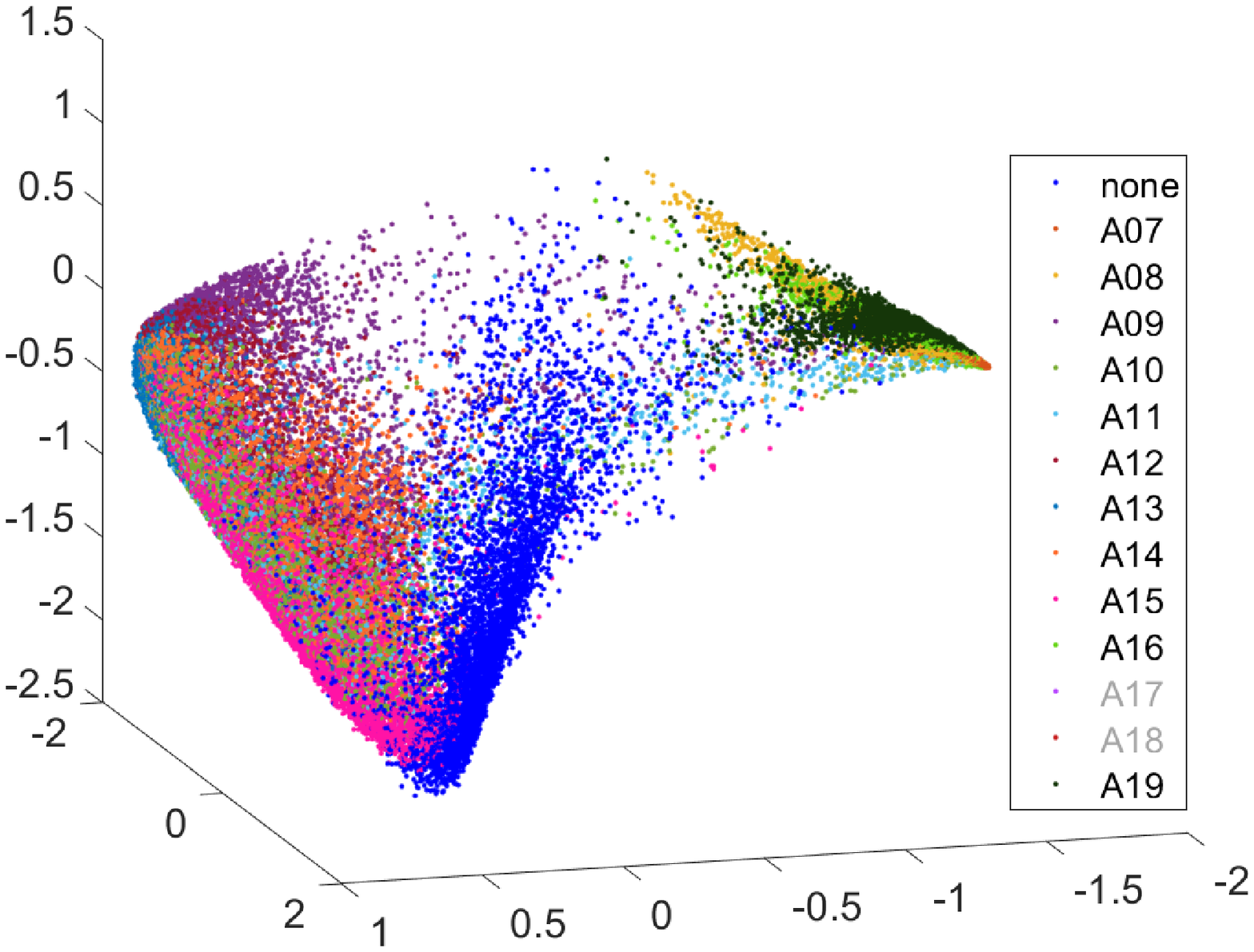}
        \caption{evaluation set}
        \label{fig:allChAndDist_DM_female_eval}
    \end{subfigure}    
    \caption[DM embedding on all channels and distance metrics of Gammatone and Inverse Gammatone with 10 channels.]
    {DM embedding on all channels and distance metrics of Gammatone and Inverse Gammatone with 10 channels.
    The embeddings in this figure were made on all female samples.}
    \label{fig:allChAndDist_DM_female}
\end{figure}

In this experiment, we show how the "full filter-bank feature vector" is embedded using a DM, and how this embedding clusters different spoofing attacks. It is interesting to see in Figure~\ref{fig:gm10_vs_gmInv10_dm_all_dists_and_metrics_sub_and_norm_dev_female} how each filter-bank acts differently on the attacks. The Gammatone manages to separate well A02, A03 and A05 (all share the same waveform generator) while the Inverse Gammatone separates A01, A04 and A06. Figure~\ref{fig:allChAndDist_DM_female} demonstrates how the concatenation of the filter-banks using the DM enables full separation of all training attacks.

\subsection{Reduced Feature Embeddings}
\label{ssec:reduced_feature_embeddings}
\begin{figure}[t]
    \centering
    \includegraphics[width=0.5\linewidth]{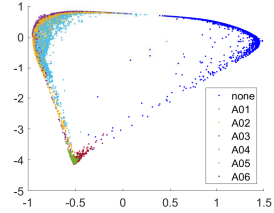}
    \caption{DM embeddings of "reduced configuration" on female development set.}    
    \label{fig:reduced_configuration_embedding_dev_female}
\end{figure}

Not all components of channels and metrics contribute to the separative characteristics of the "full feature vector". Removing some of them affects the geometric layout of the embedded samples, and in some cases increases the linear separability of the clusters. An example for such a configuration is keeping only the 5 low channels from each filter-bank, and selecting the similarity measures as follow: measures 1-5 for Gammatone and measures 1-6 excluding 4 for the Inverse Gammatone (the numbers correspond to the order by which the similarity measures were presented in Section~\ref{sec:feature_extraction}). We will refer to this configuration as the "reduced configuration".
An illustration of the embeddings under "reduced configuration" is given in Figure~\ref{fig:reduced_configuration_embedding_dev_female} and its error rate table using a LR is given in Table~\ref{tab:error_rates_reduced_configuration}. The front-ends are the 4 dimensional DM embeddings, and the threshold of the LR is set according to EER of each set.
The table shows how A05 acts differently according to the speakers' gender and that A17 and A18 (both are voice conversion attack) are the hardest attacks across all seen and unseen attacks.

\begin{table}
    \centering
    \caption[Error rate according to EER threshold under "reduced configuration".]
    { Error rate according to EER threshold under "reduced configuration".
    The error rates are given in [female] / [male] format. The overall EER is the same as the "None"}
    \begin{adjustbox}{max width=\linewidth}
        \begin{tabular}{lll}
            Attack & Error on Train [\%] & Error on Dev [\%] \\ 
            \hline 
            None & 0.33 / 5.11 & 0.12 / 2.07 \\ 
            A01 & 0.00 / 0.00  & 0.00 / 0.00 \\ 
            A02 & 0.00 / 0.06  & 0.00 / 0.00 \\ 
            A03 & 0.00 / 0.12  & 0.08 / 0.00 \\ 
            A04 & 0.00 / 0.00  & 0.00 / 0.00 \\ 
            A05 & 0.58 / 30.48 & 0.48 / 12.58 \\ 
            A06 & 1.61 / 0.00  & 0.16 / 0.00 \\ 
            \hline 
        \end{tabular}
        \begin{tabular}{ll}
            Attack & Error on Eval [\%] \\ 
            \hline 
            None & 12.99 / 12.09 \\ 
            A07 & 0.00 / 0.00 \\ 
            A08 & 0.00 / 0.00 \\ 
            A09 & 0.00 / 1.78 \\ 
            A10 & 2.76 / 2.05 \\ 
            A11 & 1.88 / 2.97 \\ 
            A12 & 0.00 / 0.13 \\ 
            A13 & 0.00 / 0.06 \\ 
            A14 & 0.61 / 1.25 \\ 
            A15 & 11.46 / 4.16 \\ 
            A16 & 0.00 / 0.00 \\ 
            A17 & 74.01 / 67.72 \\ 
            A18 & 78.22 / 77.44 \\ 
            A19 & 0.00 / 0.00\\ 
            \hline 
        \end{tabular}
    \end{adjustbox}
    \label{tab:error_rates_reduced_configuration}    
\end{table}

In Table~\ref{tab:databases_compare_reduced_config} we present the total EER by gender across two databases, ASVspoof2015 and ASVspoof2019, using the "reduced configuration". We have used 4 DM features for male samples, and 5 for female (the number of features was chosen according the development sets of both databases).
It is obvious that the advanced spoofing techniques presented in ASVspoof2019 make it harder to find a generalizing spoofing countermeasures.

\begin{table}[H]
    \centering
    \caption[EER comparison on "reduced configuration" representation across different LA databases.]
    {EER comparison on "reduced configuration" representation across different LA databases. The results were calculated using logistic regression and are given by set: train / dev / eval.}
    \begin{adjustbox}{max width=\textwidth}
        \begin{tabular}{lll}
            Database & Female EER [\%] & Male EER [\%] \\
            \hline 
            ASVspoof2015 & 0.57 / 0.35 / 1.83 & 1.40 / 2.80 / 2.27 \\ 
            ASVspoof2019 & 0.32 / 0.12 / 12.99 & 5.11 / 2.07 / 12.09 \\
            \hline 
        \end{tabular}
    \end{adjustbox}
    \label{tab:databases_compare_reduced_config}    
\end{table}


\section{Conclusions}

In this work we have shown a new time-domain PMF-based perspective to represent genuine and spoofed audio.
We have utilized diffusion maps for both visualization and feature reduction.
The use of the DM has helped us to design a feature extraction scheme and to better understand the role of each feature.
We showed that under our representation the separative information lies mostly in low to mid filter-bank channels and that some spoofing attacks acts differently on different genders.
In addition, the use of DM embedding allowed us to use a very simple linear classifier for gaining decent error rates.

\bibliographystyle{IEEEtran}
\bibliography{mybib}

\end{document}